\documentclass{icrc}

\usepackage{times}
\usepackage{amsmath}
\usepackage{amssymb}

\usepackage[dvips]{graphicx} 

\title{Propagation of cosmic ray electrons in the Galaxy}
\author{A.A. Lagutin}
\author{K.I. Osadchiy}
\author{D.V. Strelnikov}

\affil{Altai State University, Barnaul 656099, Russia}

\correspondence{Strelnikov (sdv@theory.dcn-asu.ru)}

\firstpage{1} \pubyear{2001}

\newcommand{\rr}{\vec{ r}}

\newcommand{\sphs}[3]{q_{#1}^{(#2)}\left(#3\right)}

\newcommand{\fPsi}[3]{\Psi_{#1}^{(#2)}\left(#3\right)}

\begin{document}
\maketitle
\begin{abstract}
We have made a new calculation of the cosmic ray electron spectrum
using an anomalous diffusion model to describe the propagation of
electrons in the Galaxy. The parameters defining the anomalous
diffusion have been recently determined from the study of nuclei
propagation. The predicted electron spectrum is in a good
agreement with the measurements. The source spectral index, found
from experimental data, in this approach turns out to be equal to
2.95.
\end{abstract}



\section{Introduction}
Observations of non-thermal radiation of the Galaxy stimulated
investigations of propagation of cosmic ray electrons through the
interstellar medium. Since basic paper of \citet{as1}, the problem
of calculation of electron spectrum was considered in series of papers (see, for example, \citet{as2,as3,as4,%
as5,as7,as8,Berezinski et al.(1990),Atoyan et
al.(1995),Mosk&Strong}). The normal diffusion equation for
concentration of the electrons with energy $E$,
\\$N(\vec{r},t,E)$, generated by sources distribution with density
function $S(\vec{r},t,E)$,
\begin{multline}\label{diff eq 1}
\frac{\partial N}{\partial t}=D\Delta
N(\rr,t,E)+\frac{\partial}{\partial
E}\bigl(b(E)N(\rr,t,E)\bigr)\\+S(\rr,t,E),
\end{multline}
has been used to study the electron energy spectrum modifications
in the interstellar medium (ISM). In the equation (\ref{diff eq
1}) $D$ is the diffusivity, $b(E)$ --- the energy-loss rate of
electrons.

Recently, in the papers (\citet{LagNikUch:2001,as11,LU:this
conf.}), new view of the cosmic ray propagation problem was
presented. It has been shown that the ``knee'' in the primary
cosmic ray spectrum is due to large free paths (``L\'{e}vy flights
'') of cosmic rays particles between magnetic domains --- traps of
the returned type. As the ``L\'{e}vy flights '' distributed
accordingly to inverse power law $\propto Ar^{-3-\alpha}$,
$r\rightarrow \infty$, $\alpha< 2$, is an intrinsic property of
fractal structures, in the fractal-like medium the normal
diffusion equation (\ref{diff eq 1}) certainly does not hold.

Based on these arguments in \citep{LagNikUchelec:2001} an
anomalous diffusion (superdiffusion) model for describing of
electrons transport in the fractal-like ISM was proposed. This
superdiffusion equation for concentration of the electrons without
convection has been presented in the form
\begin{multline}\label{sup-diff eq 2}
\frac{\partial N}{\partial
t}=-D(E,\alpha)(-\Delta)^{\alpha/2}N(\rr,t,E)\\+\frac{\partial}{\partial
E}\bigl(b(E)N(\rr,t,E)\bigr)+S(\rr,t,E),
\end{multline}
where $D(E,\alpha)$ is the anomalous diffusivity and
$(-\Delta)^{\alpha/2}$ is the fractional Laplacian (called
``Riss'' operator, \citet{as13}).

The solution of superdiffusion equation (\ref{sup-diff eq 2}) in
the case of point impulse source with inverse power spectrum and
the behaviour of energy spectrum of electrons in high energy
region were found.

The main goal of this paper is to calculate the spectrum of
electrons from sub-GeV to TeV energies in the framework of
anomalous diffusion model . We don't use the assumption that the
mean time of particle staying in a trap is finite. In this paper,
similarly to \citep{LU:this conf.}, we suppose that a particle can
spend a long time in a trap, that is $q(t)\propto Bt^{-\beta-1},\
t\rightarrow \infty,\ \beta<1$ (``L\'{e}vy trapping time ''). The
spectrum of electrons both in ISM and in the solar system is found
too.

\section{Flux of electrons from point source}

The flux of electrons, $J(\vec{r},t,E),$ is related to the
source\\ $S(\vec{r_0},t_0 ,E_0 )$ by the propagator
$G(\vec{r},t,E;\vec{r_0},t_0 )$:
\begin{multline}\label{J(r,t,E)}
J(\vec{r},t,E)=\frac{c}{4\pi}\int
d{\vec{r_0}}\int\limits_{E}^{\infty}dE_0
\int\limits_{-\infty}^{t}dt_0\\ \times G(\vec{ r},t,E;\vec{
r_0},t_0 )S(\vec{ r_0},t_0 ,E_0 )\delta(t-t_0 -\tau ).
\end{multline}
Here
\begin{equation}\label{tau}
\tau =\int\limits_{E}^{E_0}\frac{dE^{\prime}}{b(E^{\prime})},
\end{equation}
$\delta(t-t_0 -\tau )$ reflects the law of energy conservation in
the continuous losses approach.

The propagator in the anomalous diffusion model under
consideration has the form (see, \citet{LU:this conf.})
\begin{multline}\label{Green:s}
G(\vec{
r},t,E;\vec{r_0},t_0)=\bigl(D(E,\alpha,\beta)t^\beta\bigr)^{-3/\alpha}\\
\times \fPsi3{\alpha,\beta}{|\vec{ r}|\bigl(
D(E,\alpha,\beta)t^\beta\bigr)^{-1/\alpha}},
\end{multline}
where
\begin{equation}\label{Psi}
\fPsi3{\alpha,\beta}{r}=\int\limits_0^\infty
\sphs3\alpha{r\tau^\beta} \sphs1{\beta,1}{\tau}
\tau^{3\beta/\alpha}\, d\tau.
\end{equation}
Here \(\sphs3\alpha r\) is the density of three-dimensional
spherically-symmetrical stable distribution with characteristic
exponent \(\alpha\leq2\)~(\citet{Zolotarev:1999}) and
\(\sphs1{\beta,1}{t}\) is one-sided stable distribution with
characteristic exponent \(\beta\) \citep{Uchaikin:1999}. The
parameters $\alpha,\ \beta$ are determined by the fractional
structure of ISM and by trapping mechanism correspondingly, the
anomalous diffusivity $D(E,\alpha,\beta)$ --- by the constants $A$
and $B$ in the asymptotic behaviour for  ``L\'evy flights'' ($A$)
and ``L\'evy waiting time'' ($B$) distributions:
\[
D(E,\alpha,\beta)\propto A(E,\alpha)/B(E,\beta).
\]

The energy-loss rate of relativistic electrons is described by the
equation (see \citet{Atoyan et al.(1995)})
\begin{multline}\label{loos}
-\frac{dE}{dt}=b(E)=b_0+b_1 E +b_2 E^2\\ \approx b_2(E+E_1 )(E+E_2
),
\end{multline}
where $b_0 =3.06\cdot 10^{-16}\cdot n\ (GeV\cdot s^{-1})$ is for
ionization losses of electrons in ISM with number density $n\
(cm^{-3})$, $b_1E$ with $b_1 =10^{-15}\cdot n \ (s^{-1})$
corresponds to the bremsstrahlung energy losses, and $b_2E^2$ with
$b_2 = 1.38\cdot 10^{-16}\ (GeV\cdot s)^{-1}$ --- synchrotron and
inverse Compton losses, for $ B\approx 5\mu G$ and $\omega\approx
1(eV/cm^3)$, $E_1 \approx b_0/b_1$, $E_2 \approx b_1/b_2 $. Using
(\ref{loos}), the solution of the equaion (\ref{tau}) relative to
$E$ can be presented in the form \citep{Atoyan et al.(1995)}
$$E_0(\tau) =\frac{E+E_1 }{1-(1-e^{-b_1 \tau })(E+E_2 )/(E_2 -E_1
)}-E_1.$$ Taking into account that $b_1 \tau\leq 3.15\cdot
10^{-8}\cdot n\cdot 10^5 =3.15\cdot n\cdot 10^{-3}\ll 1,$ we
derive from the later equation
\begin{equation}\label{E00}
  E_0(\tau)=\frac{E+E_1 }{1-b_1 \tau
(E+E_2 )/(E_2 -E_1 )}-E_1 .
\end{equation}

With help of the equations (\ref{J(r,t,E)}, \ref{Green:s},
\ref{E00}) it's easy to calculate the flux of electrons for the
sources interesting for astrophysics. For example, for point
impulse source $$S(\vec{ r},t,E)=S_0 E^{-p} \delta(\vec{ r})
\Theta(T-t)\Theta(t),$$
\[
\quad \Theta(x)=\left\{\begin{array}{ll} 1,&x>0,\\ 0,&x<0, \\
\end{array}\right.
\]
we have
\begin{multline}\label{diff-solve}
J_T(\vec{ r},t,E)=\frac{c}{4\pi}S_0
\int\limits_{\max[0,t-T]}^{\min\left(t,{1}/{b_2 (E+E_2 )}\right)}
d\tau {E_0 (\tau )}^{-p}\\ \times (\lambda(E,\tau
^{\beta})\tau^{\beta-1})^{-3/\alpha} (1-b_2 \tau (E+E_2 )^{-2}\\
\times\fPsi3{\alpha,\beta}{r(\lambda(E,\tau
^{\beta})\tau^{\beta-1})^{-1/\alpha}},
\end{multline}
where $$\lambda(E,\tau )=\int\limits_{E}^{E_0
(\tau)}\frac{D(E^{\prime},\alpha,\beta)}{b(E^{\prime})}dE^{\prime}.$$

It should be noted that in the case $\beta=1$, the equation
(\ref{diff-solve}) comes to the solution, obtained earlier by
\citep{LagNikUchelec:2001}. If $\alpha=2,\ \beta=1$, we have the
standard solution.
\begin{figure*}[t]
\includegraphics[width=17.0cm]{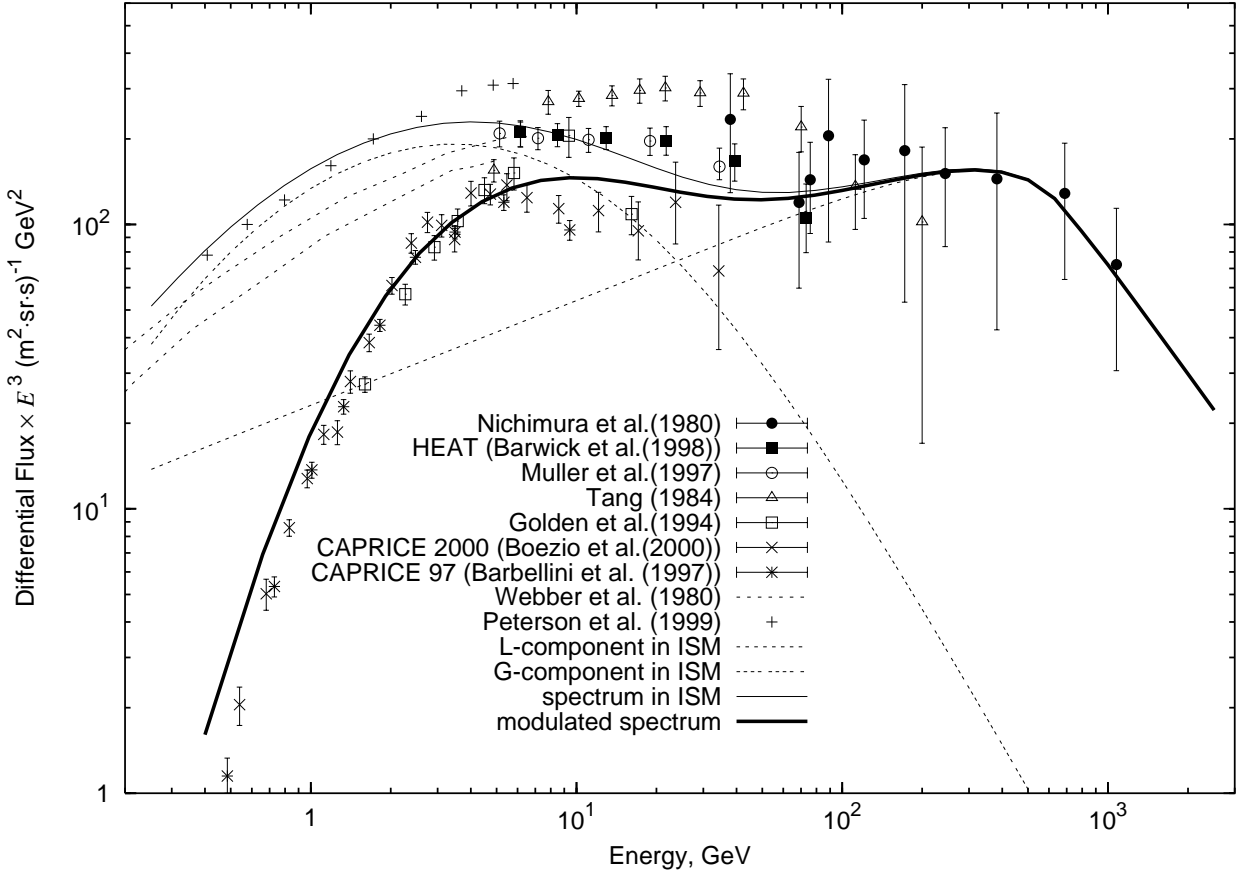}
\caption{Energy spectrum of electrons near the solar system.}
\end{figure*}
\section{Energy spectrum of electrons}

The flux $J$ of electrons due to all sources of Galaxy can be
separated into two components:
\begin{multline}\label{J_i}
  J=\int\limits_0^{1 kpc}{\ldots}+\int\limits_{1 kpc}^R{\ldots}=\\J_L(r\leq 1 kpc)+J_G(r> 1 kpc).
\end{multline}
The similar separation is frequently used in the studies of cosmic
rays (see, for example, \citet{Atoyan et al.(1995)} and references
therein).

The first component (L) in (\ref{J_i}) describes the contribution
of the nearby sources (at distance $r\leq 1 kpc$) to observed flux
$J$. The second component (G) is the contribution of the distant
sources ($r> 1 kpc$) to $J$.

The nearby sources used in our calculations is presented in
Table~1. Based on result (\ref{diff-solve}) we suppose
\begin{equation}\label{J_L}
  J_L=\sum_{\substack{i
\\ (r\leq 1
  kpc)}}{J_T(\vec{r_i},t_i,E)},
\end{equation}
where injection time $T\approx 10^4\div 10^5 y$.

The second component (G) is evaluated under assumption that the
distant sources ($r>1 kpc$) are distributed uniformly both in
space and time in the Galaxy.

The parameters defining the anomalous diffusivity and used in our
calculations have been recently derived from the study of nuclei
propagation \citep{as11}: $\alpha=1.7$; $\beta=0.8$;
$D(E,\alpha,\beta)=D_0(E/1 GeV)^\delta$ with $D_0\approx (1\div
4)\cdot 10^{-3}\ pc^{1.7}y^{-0.8}$ and $\delta=0.27$. Only one
parameter $p$ defining injection spectrum of electrons in the
sources is found by fit. Extensive calculations show that the best
fit of experimental data may be get at $p\approx 2.95$.

The spectra of L- and G- components and the total spectrum in ISM
are demonstrated in Fig. 1.

To describe the influence of the solar modulation  , the force
model of \citet{Axford} is used: $$J_{mod}(E)=\frac{E^2
-(m_ec^2)^2 }{[E+\Phi (t)]^2 -(m_ec^2)^2 }J[E+\Phi (t)],$$ where
$\Phi (t)=600 MeV.$ The modulated spectra are shown in Fig.1 too.

\section{Conclusion}

\balance We have made a new calculation of the cosmic ray electron
spectrum using an anomalous diffusion model to describe the
propagation of electrons in the Galaxy. The parameters defining
the anomalous diffusion have been recently determined from the
study of nuclei propagation. The predicted electron spectrum is in
a good agreement with the measurements. The source spectral index,
found from experimental data, in this approach turns out to be
equal to 2.95. The proximity of this exponent to one obtained
earlier~\citep{as11} for nuclei components ($p\approx2.9$) can
indicate the same mechanism of particle acceleration.

We have shown that the sources of high-energy electrons
($E\geq100$~GeV), observed in the solar system are relatively
young local sources ($r\leq 200$~pc, $t\sim10^5$~y), injecting
particles during the time $T\sim 10^4\div10^5 y$. The behaviour of
spectrum in the low-energy region is defined by distant ($r\ge1
kpc$) sources.

\begin{table}[hbt]
\caption{List of the nearest SNR}\label{tab:sources}
\begin{tabular}{lccc}
\hline \multicolumn{1}{c|}{Name}&\multicolumn{1}{|c|}{$r$, pc}&
\multicolumn{1}{c|}{$t,~10^5$~y}& \multicolumn{1}{c}{Source}\\
\hline Lopus Loop & 400 & 0.38 & \citep{as8}\\ Monoceros & 600 &
0.46 & \citep{as8}\\ Vela & 400 & 0.11 & \citep{as8}\\ Cyg. Loop &
600 & 0.35 & \citep{as8}\\ CTB 13 & 600 & 0.32 & \citep{as8}\\ S
149 & 700 & 0.43 & \citep{as8}\\ STB 72     & 700 & 0.32 &
\citep{as8}\\ CTB 1 & 900 & 0.47 & \citep{as8}\\ HB 21 & 800 &
0.23 & \citep{as8}\\ HB 9 & 800 & 0.27 & \citep{as8}\\ Monogem &
300 & 0.86 & \citep{as20}\\ Geminga    & 400 & 3.4  &
\citep{as20}\\ Loop I & 100 & 2.0 & \citep{as21}\\ Loop II & 175 &
4.0  & \citep{as21}\\ Loop III & 200 & 4.0  & \citep{as21}\\ Loop
IV & 210 & 4.0  & \citep{as21}\\
\end{tabular}
\end{table}


\end{document}